\documentclass[twocolumn, prl, showpacs, amsmath,amssymb, superscriptaddress,longbibliography]{revtex4-1}
\usepackage{graphicx}
\usepackage{dcolumn}
\usepackage{bm}
\usepackage{color}
\usepackage[T1]{fontenc}
\usepackage[colorlinks,linkcolor=blue,citecolor=blue,urlcolor=blue]{hyperref}
\usepackage{soul}
\usepackage[normalem]{ulem}
\usepackage{color}

\begin{document}

\pdfstringdefDisableCommands{\let\sout\relax}

\title{Fano interference for tailoring near-field radiative heat transfer}

\author{J. E. P\'{e}rez-Rodr\'{i}guez}
\affiliation{Instituto de F\'{i}sica, Universidad Nacional Aut\'{o}noma de M\'{e}xico, Apartado Postal 20-364, M\'{e}xico D.F. 01000, M\'{e}xico}

\author{Giuseppe Pirruccio}
\affiliation{Instituto de F\'{i}sica, Universidad Nacional Aut\'{o}noma de M\'{e}xico, Apartado Postal 20-364, M\'{e}xico D.F. 01000, M\'{e}xico}

\author{Ra\'{u}l Esquivel-Sirvent}
\email{raul@fisica.unam.mx}
\affiliation{Instituto de F\'{i}sica, Universidad Nacional Aut\'{o}noma de M\'{e}xico, Apartado Postal 20-364, M\'{e}xico D.F. 01000, M\'{e}xico}

\date{\today}

\begin{abstract}

We show the existence of Fano resonances in the context of near-field radiative heat transfer, which enables the strong suppression and enhancement of the spectral heat flux at specific wavelengths. We make use of the plasmon-phonon coupling in a symmetric nano cavity composed of a polaritonic material coated with a metallic layer. Each side of the cavity is kept at different temperatures. The  hybridization of the plasmonic and phononic modes sustained by the multilayer structure is determined by the matching of their polarization. This leads to the opening of a thermal band gap, where heat transfer in the coupled system is inhibited for all wave vectors and for wavelengths at which the individual constituents are thermally transmissive.

\end{abstract}

\maketitle

Tuning and modulation of near-field radiative heat transfer (NFRHT) have recently attracted great interest due to its relevance for various technological applications such as thermal transistors \cite{Transistor}, nanophotonic thermal devices \cite{pvoltaic}, thermal diodes \cite{diodes} and heat-assisted data storage \cite{magrec}.

Several strategies have been adopted to control the NFRHT exploiting periodic gratings \cite{Gelais,PhysRevB.95.125404}, dielectric coatings \cite{Fu20091027}, graphene-coated materials \cite{plasmongraphene}, phase-change materials \cite{vanZwol,vanZwol2,Yue}, porous materials \cite{Biehsporous,Thinfilmraul,Esq17}, metamaterials \cite{Basu3,Basu} and external magnetic fields \cite{PhysRevB.92.125418}.  Silicon-based meta surfaces have been shown to provide the largest radiative heat conductance \cite{garciavidal}.

Near-field radiative heat transfer is mediated by propagating and evanescent modes between two media kept at different temperatures. The former dominate when the two media are separated by a distance larger than the thermal wavelength, while the latter gives a large contribution to the total heat flux when the two media are close enough that  near-field overlap takes place \cite{HARGREAVES,vanhove}. Near field heat transfer can largely exceed the one by propagating modes only, that corresponds to the classical prediction of Stefan-Boltzman law for black bodies \cite{Song15} and is a highly coherent phenomenon \cite{JONES2013349}. Several experiments have shown this anomalous heat transfer at the nanoscale \cite{hillenbrand,hillenbrand2,dewilde,kittel,Rousseau,Gelais,gelaisnearfield,Francoeur15,Song,songradiative,giantkittel}.

A very efficient mechanism for heat transfer is found in materials sustaining surface phonon-polaritons \cite{Joulain200559}. The enhanced heat flux found at specific wavelengths is determined by the coherent interaction of  the evanescent near-fields.  So far, researchers have focused on the phonon contributions to the heat flux in half spaces, single layers and multilayers \cite{Philthin,Bright,amplification}. However, the possibility of hybridization between different modes in a multilayer structure remains unexplored. 

Micro- and nanocavities allow the spatial confinement of different types of excitations in the same volume. For instance, the coexistence of mechanical vibrations and optical waves in a micro resonator gave birth to the field of optomechanics \cite{Aspelmeyer}. Moreover, because of the ever shrinking of electronic devices, they are believed to hold a role of primary importance in next generation electronics. An important issue is the control of the local heating and cooling, which is critical to maintaining high performances and ensure stability. The archetype structure for a cavity consists of two homogeneous, parallel and semi-infinite volumes separated by a gap and kept at different temperatures.

In nano-optics the near-field interaction has been exploited to demonstrate enhancement and suppression of optical properties of coupled systems. The coupling of a spectrally broad resonance with a narrow one leads to a distinctive phenomenon known as Fano resonance, where regions of constructive and destructive interference appear in the spectrum \cite{fano}. Fano resonances were first observed in atomic systems, and only recently an analogous mechanism has been observed in plasmonic systems, through dramatic changes in the nanostructures scattering cross section, extinction and photoluminescence of light emitters coupled to them \cite{fano2,Revfano,giessen,weakstrong}.

In this Letter, we propose a mode hybridization scheme to tailor NFRHT. We make use of a symmetric nanocavity made of  metal-coated dielectric layers, to induce a strong coupling between surface plasmon-polaritons in the metal and surface phonon-polaritons in the dielectric. This gives  rise to Fano resonances which dramatically affect the spectral heat flux, eventually opening a large tailorable bandgap.

Within the fluctuating electrodynamics formalism \cite{Vinogradov}, the NFRHT is  described by the spectral heat
flux $S_{\omega}(T_1,T_2,d)$, expressed as \cite{Song15,Thinfilmraul}
\begin{equation}
S_{\omega}(T_1,T_2,d)=(\Theta(\omega,T_2)-\Theta(\omega,T_1))\sum_{j=p,s}\int\frac{\beta d\beta}{(2\pi)^2} \tau_j(\omega,\beta,d),
\label{flux}
\end{equation}
where  the sum runs over the two possible polarizations $p$ or $s$, $\omega$ is the frequency, $\Theta(\omega,T)=\hbar \omega(exp(\hbar \omega/k_B T)-1)^{-1}$ is the Planck distribution function, $\beta$ is the component of the wave vector parallel to the interfaces, related to the normal component $\kappa$ by the relation $\kappa^2=(\omega/c)^2-\beta^2$ in vacuum and $\kappa_i^2=(\omega/c)^2\epsilon_i-\beta^2$ in a material with dielectric function $\epsilon_i$.

  The energy transmission coefficient, $\tau_j(\omega,\beta,d)$, has two contributions, one from propagating waves ($\beta <\omega /c$) and the second one from evanescent waves ($\beta> \omega /c$),
\begin{equation}
\tau_{p,s}(\omega,\kappa,d)=
\begin{cases}
(1-|r_{p,s}|^2)^2/|1-r_{p,s}^2 e^{2i\kappa d)}|^2, & \text{if $\beta <\omega /c$} \\
4 Im(r_{p,s})^2e^{-2 |\kappa|  d}/|1-r_{p,s}^2 e^{-2|\kappa|d}|^2, & \text{if $  \beta> \omega /c$}
\end{cases}
\label{trans}
\end{equation}
where $r_{p,s}$ are the reflectivities written in terms of the effective surface impedances \cite{Impedance}  to account for the layered structures and the finite thickness of the slabs.
 \begin{figure}[b]
\centerline{\includegraphics[width=1.\columnwidth,draft=false]{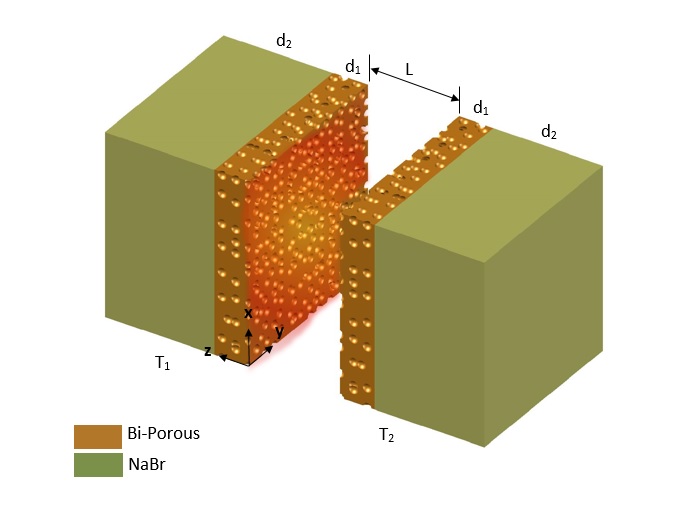}}
\caption{Geometry of the symmetric nanocavity formed by a polaritonic material (NaBr) coated by a layer of porous Bi. The two sides of the cavity are kept at different temperature ($T_1=300$ K and $T_2=0$ K).}
\label{schematics}
\end{figure}

The system we consider, shown in Figure(\ref{schematics}), is composed of two metal-coated dielectric layers separated by a gap of $L$=10 nm.  Each dielectric layer is made of a polaritonic material, NaBr, while the metallic coating is made of porous Bi.  The thickness of  Bi is $d_1$=75 nm, while the thickness $d_2$ of  NaBr varies between 75 and 750 nm. The temperature difference within the cavity is  $300$ K, well below the Wien wavelength associated to $T_1$ and $T_2$. The dielectric function of the NaBr is a Lorentz type $\epsilon_{NaBr}= 2.6(\omega^2-\omega_{LO}^2+i\omega \gamma_a)/(\omega^2-\omega_{TO}^2+i\omega \gamma_a)$, where $\omega_{LO}=39\times10^{12}$ rad/s is the longitudinal phonon frequency, $\omega_{TO}=25\times10^{12}$ rad/s is the transverse phonon frequency, and $\gamma_a=10^{-2}\omega_{T0}$ is the damping parameter \cite{Marquez}. For bulk Bi, we consider a Drude model with a plasma frequency of $\omega_p=60.77\times10^{12}$ rad/s, plus a sum of nine Lorentz oscillators \cite{opticalbi}. The dielectric function of a porous Bi layer is described by a Bruggeman effective medium approximation \cite{Esq17} in which the resulting effective plasma frequency is given by $\omega_{peff}=\omega_p\sqrt{(2-3f)/3}$, where $f$ is the porosity of the Bi \cite{opticalprops,porousplasmon}.

\begin{figure}[hh]
\centerline{\includegraphics[width=1.0\columnwidth,draft=false]{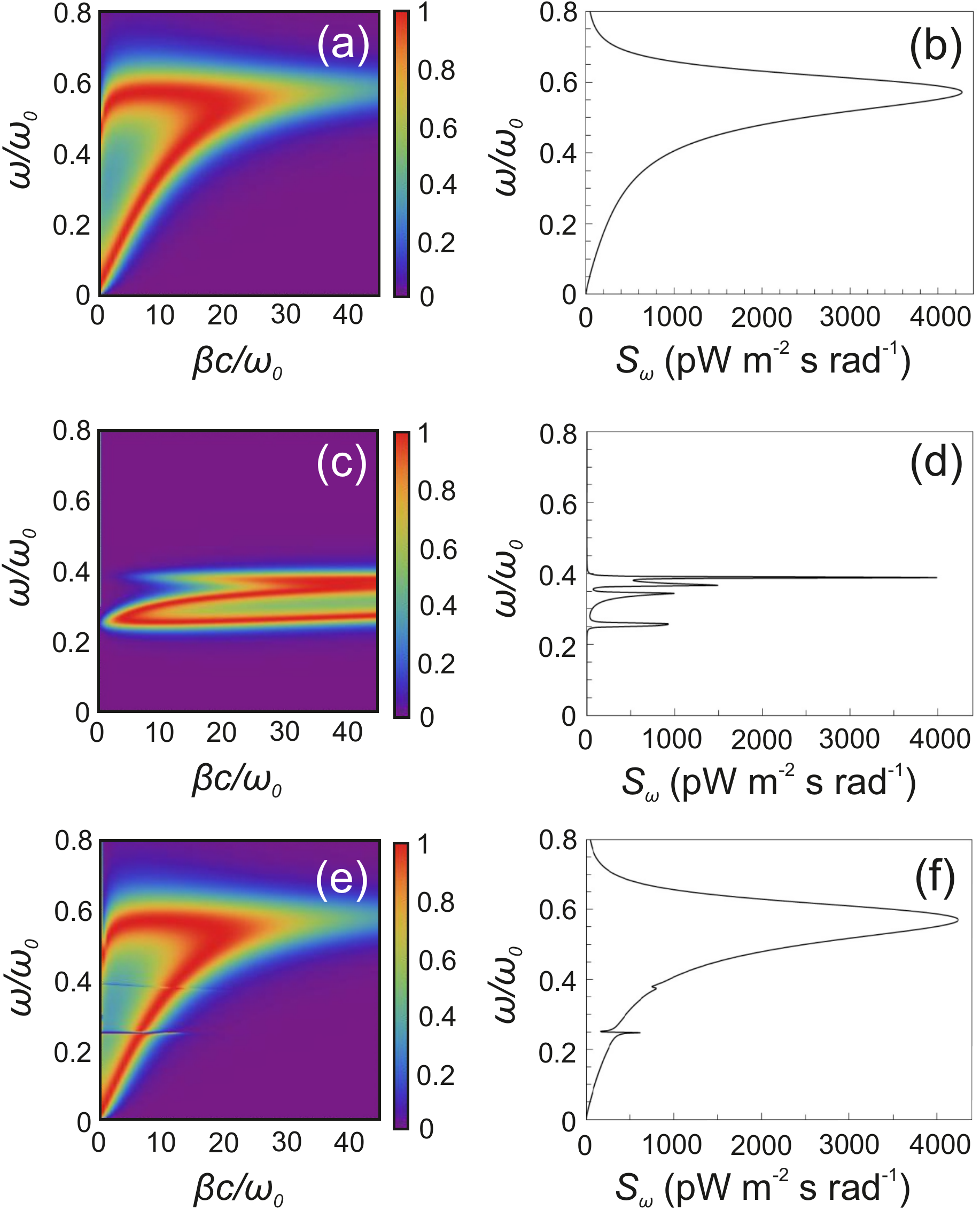}}
\caption{(a,c,e) $p$-polarized transmission coefficient as a function of the normalized frequency and normalized wave vector. (b,d,e) Spectral heat flux as a function of the normalized frequency. The cavity is formed by (a,b) two 75 nm-thick, non-porous Bi layers, (c,d) two 75 nm-thick NaBr layers and (e,f) two 75 nm-thick, non-porous Bi layers and two 75 nm-thick NaBr layers.}
\label{figure2}
\end{figure}

We first analyze the individual contributions to the NFRHT of the two NaBr layers and the two Bi layers. The left column of Fig.~\ref{figure2} displays the $p$-polarized transmission coefficient, $\tau_p$ (Eq. (\ref{trans})), as a function of the normalized frequency, $\omega/\omega_0$ with $\omega_0=10^{14}$ rad/s, and normalized wave vector $\beta c/\omega_0$. On the right column, we show the spectral heat flux, $S_{\omega}$ (Eq. (\ref{flux})), as a function of the normalized frequency.  Figures~\ref{figure2}(a-b) and (c-d) correspond to the  two non-porous 75 nm-thick Bi layers and the two 75 nm-thick NaBr layers, respectively. The light cone can be appreciated in panels (a,c,e) for values of the wave vector very close to 0. The dispersive and spectrally broad feature in Fig.~\ref{figure2}(a) is associated with the excitation of the evanescent gap surface plasmon polariton (G-SPP) \cite{nanoscale}. This mode arises from the coupling of surface plasmon polaritons sustained by the two metallic layers (see Supplemental material (SM) \cite{supplemental}) and for large values of $\beta c/\omega_0$ it approaches the value $\omega_{peff}\simeq \omega_p$. The contribution of the G-SPP to $S_{\omega}$ is shown in Fig.~\ref{figure2}(b) which is peaked in correspondence to the dispersionless regions of $\tau_p$. Figure~\ref{figure2}(c) and (d) display several sharp features corresponding to the excitation of  surface phonon polaritons (SPhPs). In the bulk case, the longitudinal and the degenerate transverse optical phonon appear for $\omega/\omega_0$=0.4 and 0.25, respectively (see SM \cite{supplemental}). In contrast,  multiple interfaces allows the excitation of several longitudinal and transverse modes. These sharp non-dispersive resonances appear in the same wavelength range of the dispersive G-SPP. Therefore, the NFRHT of the structure composed of the polaritonic material and the plasmonic one is expected to arise from the coupling of G-SPP and SPhPs, which we termed bare modes of the structure. In Fig.~\ref{figure2}(e) and (f) $\tau_p$ and $S_{\omega}$ for the complete structure are shown, respectively. The small asymmetric features around $\omega/\omega_0$=0.25 and 0.4 are due to the coupling between G-SPP and SPhPs, allowed by the longitudinal and transverse polarization components of the G-SPP, which interact with the longitudinal and transverse SPhPs, respectively. Even though a modification of $S_{\omega}$ is visible for frequencies close to the phonon ones, the overall plasmon-phonon coupling is weak because of the limited spectral overlap of the bare modes of the system.

\begin{figure}[t]
\centerline{\includegraphics[width=1.\columnwidth,draft=false]{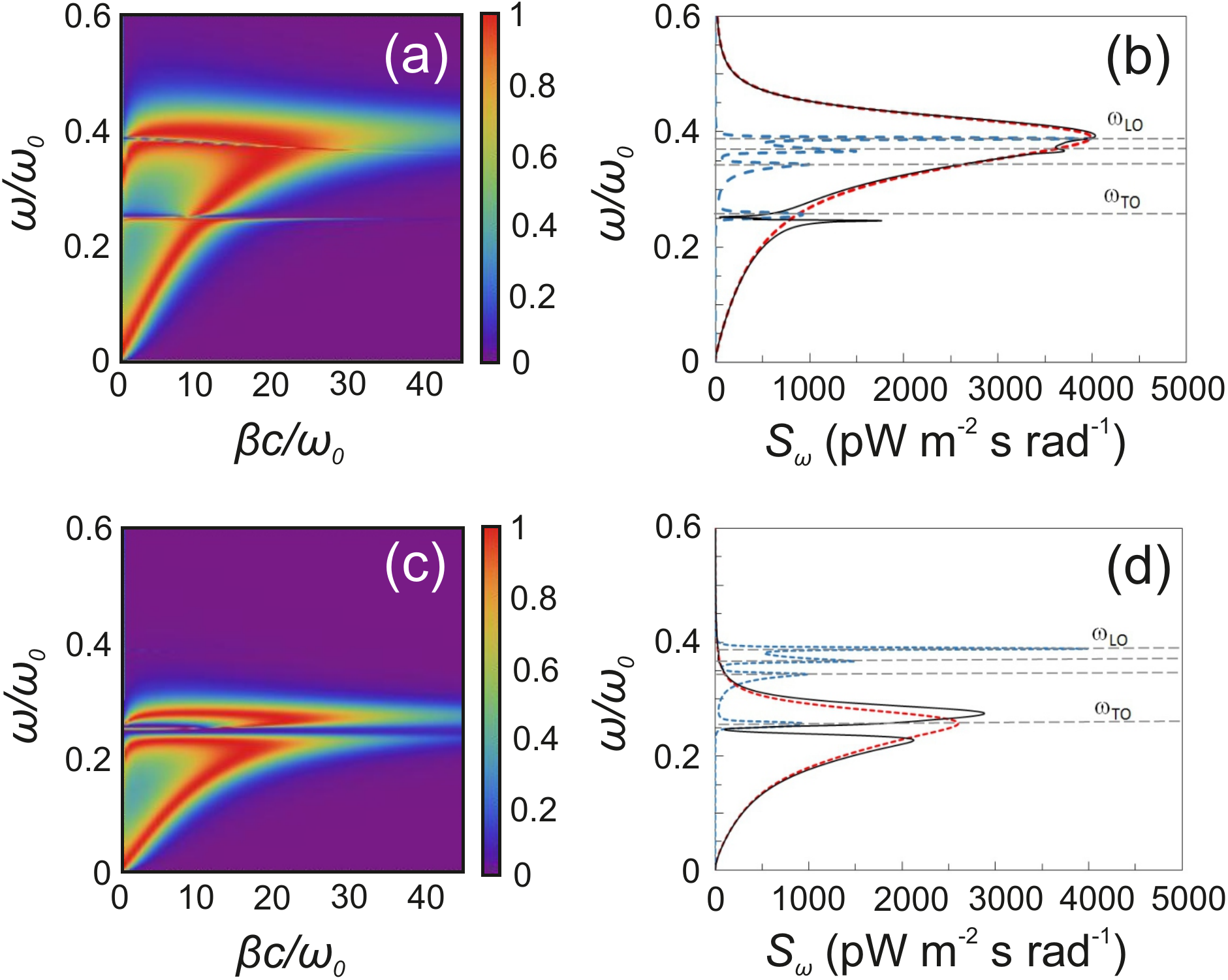}}
\caption{$p$-polarized transmission coefficient as a function of the normalized frequency and normalized wave vector for the structure displayed in Fig. 1 with a Bi porosity of (a) 0.35 and (c) 0.53. Spectral heat flux as a function of the normalized frequency for the structure displayed in Fig. 1 with a Bi porosity of (b) 0.35 and (d) 0.53. The dashed red curve represents the heat flux associated to the G-SPP; the  dashed blue curve represents  the heat flux associated to the SPhPs; and the black continuous curve represents the total heat flux of the structure.  Gray-dashed  lines indicate the position of the  longitudinal and transverse SPhPs.}
\label{figure3}
\end{figure}

 To optimize the coupling of the G-SPP and the SPhPs we vary the porosity of the Bi layer which effectively lowers its plasma frequency (see SM \cite{supplemental}). We assume that the porosity only affects the plasma frequency without introducing additional scattering channels for the bare modes. In Figs.~\ref{figure3}(a) and (c) we plot $\tau_p$ for two values of the porosity, $f$=0.35 and 0.53, which shift $\omega_{peff}$ to the frequencies of the longitudinal and transverse SPhPs, respectively. 
 
 In Fig.~\ref{figure3}(a) the plasma frequency is tuned to $\omega_{LO}$. We observe a weak coupling between the bare modes for frequencies close to $\omega_{LO}$, while close to the transverse phonon frequency, $\omega_{TO}$, a stronger coupling appears.  
 In order to analyze this mode coupling, in Fig.~\ref{figure3}(b) the contributions to $S_{\omega}$ from the NaBr layers and porous Bi layers are plotted with dashed blue and red curves, respectively. The black continuous curve represents $S_{\omega}$ for the complete structure. The hybridization between the bare modes becomes evident close to $\omega_{T0}$ where Fano interference occurs and $S_{\omega}$ displays an asymmetric line shape. This is due to the interference between the spectrally broad G-SPP and the spectrally sharp SPhPs which allows tailoring the radiative heat transfer by creating spectral regions of enhanced and suppressed heat flux with respect to the one due to the bare modes. 

By tuning the G-SPP peak to the transverse SPhP (see Figs.~\ref{figure3}(c-d)) we obtain a spectral window where the heat transfer is forbidden for all wave vectors, although $S_{\omega}$ associated to the bare modes exhibits a maximum. This phenomenon is the thermal analogy of the electromagnetically induced transparency observed in resonant photonic nanostructures \cite{imamoglu,imamoglu2,zhang08,tassin09,tassin12}. We associate the splitting of the G-SPP band to the strong coupling between the G-SPP and the transverse SPhPs resulting in the excitation of plasmon-phonon polaritons. The contribution to the NFRHT due to the coupling of the G-SPP with the longitudinal phonons is always weak or absent due to the polarization mismatch between the bare modes. Calculation for the total $S_{\omega}$  is found in the SM \cite{supplemental}.

\begin{figure}[t]
\centerline{\includegraphics[width=1.\columnwidth,draft=false]{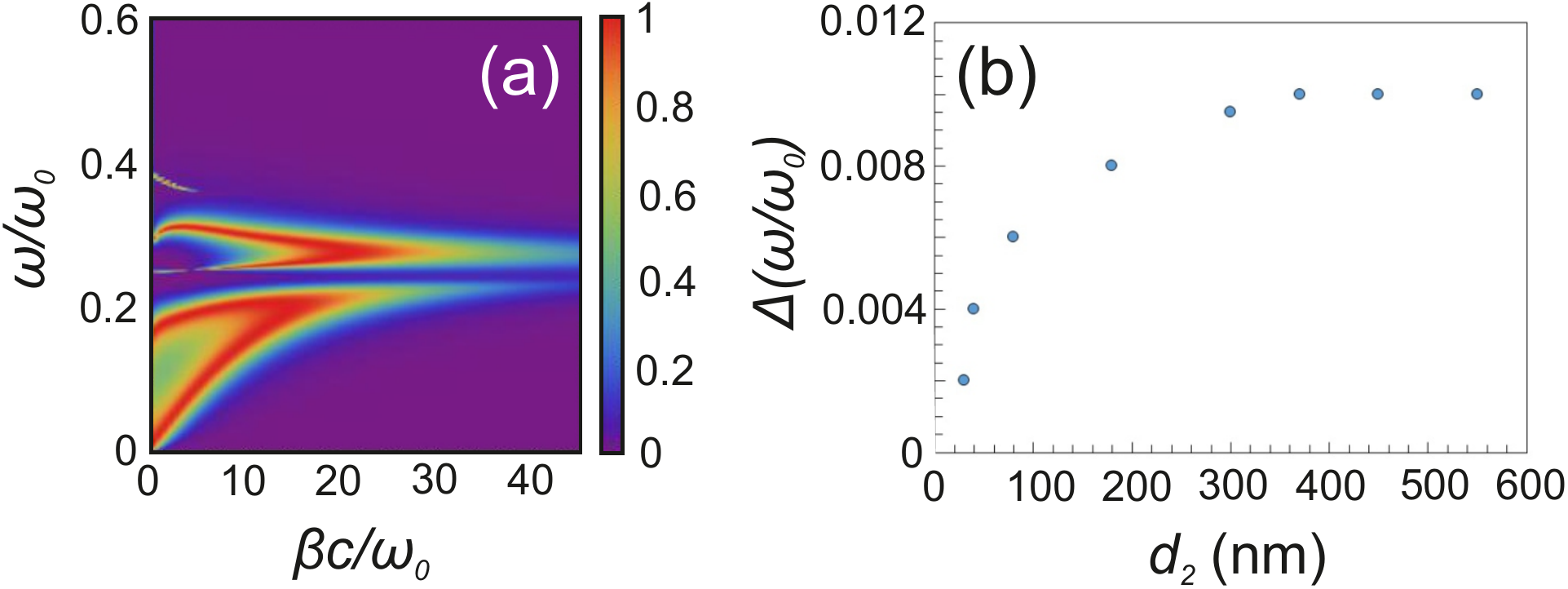}}
\caption{(a) $p$-polarized transmission coefficient as a function of the normalized frequency and normalized wave vector for the structure displayed in Fig. 1 with the thickness of the NaBr layer equals to 750 nm. (b) Width of the bandgap at $\beta c /\omega_0$=18 as a function the thickness of the NaBr layer.}
\label{figure4}
\end{figure}

The spectral window of forbidden NFRHT is determined by the coupling efficiency of the G-SPP with the transverse SPhP, which, in turn, is related to the geometry of the cavity. In Fig.~\ref{figure4} we study the opening of the thermal bandgap for different values of the thickness, $d_2$ of the NaBr layers and for $f$=0.53, while all the other geometrical parameters are kept constant. In Fig.~\ref{figure4}(a) we plot $\tau_p$ for $d_2$=750 nm, while in Fig.~\ref{figure4}(b) the width of the bandgap at $\beta c /\omega_0$=18 is plotted as a function $d_2$. We see that the width of the bandgap saturates as $d_2$ grows. This can be explained by looking at the decay length of the electric field of the plasmon-phonon polariton, which can be estimated as \cite{nanoscale} $L_d\simeq|\kappa_{NaBr}|^{-1}$  and is equal to 160 nm for $\beta c/\omega$=18. For $d\ll L_d$ the field leaks out the structure and, correspondingly, more  phonon peaks appear in $S_\omega$ (see Fig.~\ref{figure3}) with respect to the bulk case (see SM \cite{supplemental}). 

In conclusion, we demonstrated the excitation of Fano resonances and the modulation of NFRHT in a symmetric nanocavity composed of a polaritonic material coated with a metallic layer. These resonances arise from the coupling of surface plasmon polaritons and surface phonon polaritons sustained by each layer and  strong suppression and enhancement of the total spectral heat flux in specific frequency regions is shown. The mode hybridization is explained in terms of polarization matching and it causes the opening of a thermal bandgap. We show that the width of the bandgap can be controlled through the geometrical parameters of the nanocavity, 
and that under resonant conditions, the thermal properties of the nanocavity are radically different from those of its constituents. \\

This work was supported  by UNAM-DGAPA (PAPIIT IA102117, IN110916), CONACYT-postdoctoral fund 291053 and PIIF-UNAM.  We thank E. Y. Santiago for her suggestions and careful reading of the manuscript.  \\

G. P. and J. E. P. R. contributed equally to this work.

\paragraph{Supporting Information Available}
%

\end{document}